\documentstyle[aps,twocolumn]{revtex}

\begin{document}
\title{The Superconductor-Insulator Transition in 2D }
\author{N. Markovi\'{c}\thanks{%
Present address: Department of Applied Physics, Delft University of
Technology, Lorentzweg 1, 2628 CJ Delft, The Netherlands}, C. Christiansen,
A. M. Mack, W. H. Huber, and A. M. Goldman}
\address{School of Physics and Astronomy, University of Minnesota, \\
Minneapolis,\\
MN 55455, USA}
\date{March 4, 1999}
\maketitle

\begin{abstract}
The superconductor-insulator transition of ultrathin films of bismuth, grown
on liquid helium cooled substrates, has been studied. The transition was
tuned by changing both film thickness and perpendicular magnetic field.
Assuming that the transition is controlled by a $T=0$ critical point, a
finite size scaling analysis was carried out to determine the correlation
length exponent $\nu $ and the dynamical critical exponent $z$. The phase
diagram and the critical resistance have been studied as a function of film
thickness and magnetic field. The results are discussed in terms of bosonic
models of the superconductor-insulator transition, as well as the
percolation models which predict finite dissipation at $T=0$.
\end{abstract}




\section{INTRODUCTION}

After about two decades of research, the superconductor-insulator (SI)
transition in disordered films of metals remains a controversial subject,
mainly due to contradictory results in both theoretical and experimental
studies. This work aims to improve the understanding of this phenomenon,
which might also be relevant for high-T$_{c}$ superconductors and possibly
connected to novel metal-insulator transitions in 2D electron systems.

The superconductor-insulator transition in ultrathin films of metals is
believed to occur at the absolute zero of temperature when the quantum
ground state of the system is changed by tuning disorder, film thickness,
carrier concentration or magnetic field. Unlike finite temperature phase
transitions in which thermal fluctuations are crucial, $T=0$ phase
transitions are driven purely by quantum fluctuations. At finite
temperatures, an underlying quantum phase transition manifests itself in the
scaling behavior of the resistance with the appropriate tuning parameter and
the temperature, along with the coherence length and dynamical critical
exponents, $\nu $ and $z$ respectively\cite{Sondhi}. Various models of the
superconductor-insulator transition in disordered films can be roughly
divided in two groups: those in which the superconductivity is destroyed by
fluctuations of the amplitude of the order parameter, and those which focus
only on the phase fluctuations.

If superconductivity is destroyed only by phase fluctuations, then Cooper
pairs persist on the insulating side of the transition and the transition
may be described by a model of interacting bosons in the presence of
disorder. Based on this assumption, Fisher and co-workers \cite{Fisher g}
suggested a scaling theory and a phase diagram for a two-dimensional system
as a function of temperature, disorder, and magnetic field\cite{Fisher
B,Fisher d}. The superconducting phase is considered to be a condensate of
Cooper pairs with localized vortices, and the insulating phase is a
condensate of vortices with localized Cooper pairs. At the transition, both
vortices and Cooper pairs are mobile as they exchange their roles, which
leads to a finite resistance. Some important predictions of the model are
the universal value of this critical resistance and specific values of the
critical exponents $\nu $ and $z$.

This so-called ''dirty boson'' problem has been extensively studied using
quantum Monte Carlo simulations \cite
{Scalettar,Makivic,Cha1,Cha2,Wallin,Kisker,Wagenblast,Lidmar}, real-space
renormalization group techniques \cite{Singh,Zhang}, strong-coupling
expansions \cite{Freericks} and in other ways \cite
{Gold,Ma,Fradkin,Herbut1,Ghosal}. Finite temperature behavior in the
vicinity of a quantum critical point was also studied analytically\cite
{Sachdev,Herbut2}. A transition from a superfluid to a Mott insulator was
found in the pure case, and to a Bose glass insulator in the presence of
disorder, but there is still considerable disagreement as to the
universality class of the transition, as well as the value of the critical
resistance.

An alternative picture of interacting electrons \cite
{Maekawa,Belitz,Finkelshtein} proposes a different mechanism: the density of
states and the Cooper pairing are suppressed on the insulating side of the
superconductor-insulator transition due to an enhanced Coulomb interaction.
The SI transition occurs as a consequence of fluctuations in the amplitude,
rather than the phase of the order parameter. In other words, Cooper pairs
break up into single electrons at the transition. Therefore the
superconducting gap would also vanish at the transition.

The model of interacting electrons has also been studied numerically.
Quantum Monte Carlo simulations of an attractive fermion Hubbard model with
on-site interactions \cite{Trivedi} yielded a direct
superconductor-to-insulator transition in two dimensions without an
intervening metallic phase. The critical resistance was found to depend on
the strength of the attractive interaction, as a function of which, a
crossover from a fermionic to a bosonic regime occurs. The results of this
theory qualitatively resemble the experimental data. A recent calculation of
the effect of disorder on the gap in the density of states, using a similar
model \cite{Huscroft}, showed that the existence of a gap on the insulating
side of the transition depends on the coupling strength, allowing for a
Fermi insulator at weak and a Bose insulator at strong coupling.

Experimentally, the destruction of superconductivity by disorder has been
studied in films of MoGe \cite{Graybeal,Yazdani}, InO$_{x}$ \cite
{Hebard,Paalanen,Gantmakher1,Gantmakher2,Okuma} and Bi, Pb, Ga, Al \cite
{Jaeger,Liu,Wu} among others. Evidence was found of $T_{c}$ going to zero
with increasing disorder \cite{Graybeal} implying the destruction of Cooper
pairs at the transition. Tunneling experiments also seem to support the
fermionic picture. Valles {\it et al}. found that the superconducting gap
and the mean field transition temperature are both suppressed as disorder is
increased, and that the gap vanishes on the insulating side of the
superconductor-insulator transition \cite{Valles}. Hsu {\it et al}. carried
out tunneling studies of the superconductor-insulator transition in PbBi/Ge
films, and found a large number of quasiparticle states near the Fermi
energy \cite{Hsu}. They estimated the average number of Cooper pairs in a
coherence volume to be on the order of one at the superconductor-insulator
transition. This result, in combination with the disappearance of the energy
gap, was interpreted as evidence of the superconductor-insulator transition
being driven by fluctuations in the amplitude of the order parameter.
Alternatively, it is possible for the superconducting energy gap to be
reduced or the tunneling density of states to be broadened as a consequence
of phase fluctuations \cite{Ferrel}. Thus, the absence of the gap in these
tunneling studies does not necessarily mean that Cooper pairs are absent on
the insulating side of the superconductor-insulator transition, but it may
imply that a full picture might have to include fermionic degrees of freedom.

Evidence of the importance of the bosonic picture can be found in the work
of Paalanen{\it \ et al}.\cite{Paalanen}. These workers studied the
magnetoresistance and the Hall effect in amorphous InO$_{x}$ films and
observed two distinct transitions: one at a critical field $B_{xx}^{c}$
where the longitudinal resistance diverges and the system presumably
undergoes a transition from a superconducting phase to a Bose glass
insulator with localized Cooper pairs, and the other at a higher field $%
B_{xy}^{c}$, where the transverse resistance diverges and the Cooper pairs
of the Bose glass insulator presumably unbind. The transition in the
transverse resistance occurred at the same magnetic field where the
longitudinal resistance showed a maximum. Since a Bose insulator might be
expected to have a higher resistance than an insulator with localized single
electrons, and from the disorder dependence of $B_{xx}^{c}/B_{xy}^{c}$, this
was interpreted as evidence of the bosonic nature of the insulating state
close to the superconductor-insulator transition. Similar behavior was
observed by other groups \cite{Okuma}. Magnetoresistance studies of
amorphous $InO_{x}$ films by Gantmakher et al.\cite{Gantmakher1} also seem
to support the bosonic picture. Furthermore, a linear component of the
magnetoresistance observed in the insulating regime in amorphous Bi films
can be interpreted as a signature of vortex motion \cite{Markovic1}.

In the context of the scaling behavior, the thickness tuned transition of
ultrathin films of amorphous $Bi$ has been studied in zero magnetic field%
\cite{Liu}. A scaling analysis of the magnetic field tuned SI transition has
been carried out for thin films of $In0_{x}$ \cite{Hebard} and $MoGe$. \cite
{Yazdani} All of these investigations found $\nu \approx 1.3$ and $z\approx
1 $, consistent with the theoretical predictions of the boson Hubbard model.

Yet another interpretation of the experimental data has recently been
proposed by Shimshoni {\it et al}. \cite{Shimshoni} and expanded upon by
Mason and Kapitulnik \cite{Mason}. In this picture, a film contains both
insulating and superconducting puddles, and transport is dominated by
tunneling or activated hopping between them. The SI transition then occurs
as a consequence of the percolation of one phase or the other. Since the
correlation length exponent in 2D classical percolation is 4/3, this is
consistent with $\nu {\approx }1.3$ observed in most experiments. This model
also predicts a saturation of the resistance at very low temperatures, which
seems to be supported by the experimental data of Ephron{\it \ et al}. \cite
{Ephron}, and Yazdani and Kapitulnik\cite{Yazdani}. Similar effects have
been observed in the much earlier work of Wang {\it et al. }\cite{WangT}{\it %
\ }on underdoped high-$T_{c}$ (cuprate) films. These ideas may be relevant
to similar features of the results of Kravchenko {\it et al}. on two
dimensional electron gas systems \cite{Kravchenko}. In all studies in which
there is flattening in $R(T)$ at low temperatures, one must be concerned
with the possibility of electrical noise being the source of the effect.
Also in multi-component materials such as MoGe and underdoped cuprates there
is always a possibility of second phases affecting the outcome. Furthermore,
it has recently been proposed that the flattening in $R(T)$ at low
temperatures may be a signature of Bose metal, a phase in which the Cooper
pairs are mobile but do not condense \cite{Das1}.

The quantitative results of the study of the magnetic field tuned
superconductor-insulator transition presented here for disordered metal
systems are in serious disagreement with previous measurements of this
transition, adding yet another puzzle to this problem, and calling for a
re-examination of existing models. The thickness-tuned transition has also
been studied in a nonzero magnetic field. This allows for the construction
of a phase diagram and a direct comparison of the two different ways of
tuning the SI transition, by varying thickness or magnetic field.

This paper is organized as follows: the finite-size scaling procedures used
to determine the critical exponents are described in Section II.
Experimental details are given in Section III. Section IV focuses on the
magnetic field-tuned transition, while the analysis of the thickness-tuned
transition in finite magnetic field, which has not been studied before, is
presented in Section V. In Section VI, the phase diagram as a function of
thickness and magnetic field is presented. The critical resistance and its
apparent non-universality are discussed in Section VII. The results and
their implications are summarized and further discussed in Section VIII. A
brief account of a portion of this work has been previously reported \cite
{Markovic2}.

\bigskip

\section{SCALING\ PROCEDURES}

The scale of fluctuations on either side of a quantum phase transition is
set by a diverging correlation length $\xi \propto \delta ^{-\nu }$ and a
vanishing characteristic frequency $\Omega \propto \xi ^{-z}.$ Here $\delta $
is the deviation from the critical point $\delta =|K-K_{c}|,$ where $K$ is
the control or tuning parameter, which drives the system through the
transition (i.e. disorder, thickness, magnetic field, etc.), $K_{c}$ is the
critical value of $K$ at the transition, $\nu $ is the correlation length
exponent and $z$ is the dynamical critical exponent. The exponents $\nu $
and $z$ determine the universality class of the transition. They may not
depend on the microscopic details of the physics of the system under study,
but on its dimensionality, the symmetry group of its Hamiltonian and the
range of interactions.

The resistance of a two-dimensional system in the quantum critical regime
follows the scaling relation \cite{Sondhi,Fisher B}:

\begin{equation}
R(\delta ,T)=R_{c}\ f(\delta T^{-1/\nu z})  \label{T scaling}
\end{equation}
Here $\delta =|d-d_{c}|$ in the case of the thickness-tuned transition and $%
\delta =|B-B_{c}|$ in the case of the magnetic field-tuned transition. $%
R_{c} $ is the critical resistance at $\delta =0$, and $f(x)$ is a universal
scaling function such that $f(0)=1$.

The first step in the analysis of the experimental data is to determine the
critical value of the tuning parameter and plot the resistance as a function
of $\delta $. The $\delta $-axis is then re-scaled by a factor $t$:

\begin{equation}
R(\delta ,t)=R_{c}\ f(\delta t)  \label{rescaled}
\end{equation}
where the parameter $t(T)$ is determined at each temperature by performing a
numerical minimization which yields the best collapse of the data. If the
resistance really follows the scaling law (Eq. \ref{T scaling}), it is
obvious that $t(T)$ has to be a power law in temperature, $t(T)\equiv
T^{-1/\nu z}$. The exponent product $\nu z$ is then found by plotting $t(T)$
as a function of $T$ on a log-log scale, and determining the slope which is
then equal to $-1/\nu z$.

Similarly, at a constant temperature \cite{Yazdani}:

\begin{equation}
R(\delta ,E)=R_{c}\ f(\delta E^{-1/\nu (z+1)})  \label{E scaling}
\end{equation}
where E is the electric field across the sample. This time, the $\delta $%
-axis is re-scaled by a field-dependent factor, $t(E)$, which should be a
power law in electric field, $t(E)\equiv E^{-1/\nu (z+1)}$, and the exponent 
$\nu (z+1)$ can then be determined from the field dependence of the
parameter $t(E)$.

The main advantage of this scaling procedure is that it requires neither
prior knowledge of the critical exponents, nor the temperature and thickness
dependence of the resistance. The critical exponents are determined
empirically from the data, with the critical exponent product as the only
adjustable parameter, while the critical value of the tuning parameter is
determined independently. The temperature scaling determines the product $%
\nu z$, while the electric field scaling determines $\nu (z+1)$. Combining
the two results, the correlation length exponent $\nu $ and the dynamical
exponent $z$ can be determined separately.

An alternative way to determine these critical exponent products is to
evaluate a derivative of the resistance with respect to $K$ at its critical
value $K_{c}$ \cite{Hebard}:

\begin{equation}
(\partial R/\partial K)_{K_{c}}\propto R_{c}T^{-1/\nu z}\ f^{^{\prime }}(0)
\label{exponent}
\end{equation}
where $K\equiv d$ at the thickness-tuned transition and $K\equiv B$ at the
magnetic field-tuned transition, and $f^{\prime }(0)$ is a constant.
Plotting $(\partial R/\partial K)_{K_{c}}$ as a function of $T^{-1}$ on a
log-log scale should yield a straight line, with a slope equal to $1/\nu z$.
The same method can be applied to the electric field scaling to determine $%
1/\nu (z+1)$, and then $\nu $ and $z$ can be calculated from the results.

In the work described below, both scaling procedures were used to obtain the
critical exponents, in order to check their consistency. The exponents
obtained using two different methods were found to be the same, within the
experimental uncertainty.

\bigskip

\section{EXPERIMENTAL\ METHODS}

Ultrathin Bi films were evaporated on top of a $10\AA $\ thick layer of
amorphous Ge, which was pre-deposited onto either $SrTiO_{3}$ or glazed
alumina substrates. The substrate temperature was kept well below $20K$
during all depositions and all the films were grown{\it \ in situ} under UHV
conditions ($\sim $ $10^{-10}$ Torr). The film thickness was gradually
increased through successive depositions in increments of $0.1-0.2\AA $.
Resistance measurements were carried out between the depositions using a
standard DC four-probe technique, with currents up to 50 nA. A detailed
temperature dependence of the resistance in zero field and in magnetic field
was recorded at each film thickness in the temperature range between 0.14K
and 15K, where the lowest temperatures were achieved using a dilution
refrigerator. As the film thickness increased from 7\AA\ to 15\AA , the
temperature dependence of the resistance of the system changed from
insulator-like to superconductor-like at low temperatures, with no sign of
reentrant behavior typically observed in granular films \cite{Jaeger}. The
films that were superconducting in zero field were driven insulating by
applying a magnetic field of up to 12 kG perpendicular to the plane of the
sample using a superconducting split-coil magnet. The scaling procedures
described above were applied to the magnetic field-tuned transition, as well
as to the thickness-tuned transition in  both zero field and in a fixed
magnetic field.

\bigskip

\section{MAGNETIC FIELD-TUNED SI TRANSITION}

The resistance as a function of temperature for seven films with varying
degrees of disorder was studied in magnetic fields up to 12 kG applied
perpendicular to the plane of the sample. A typical temperature dependence
of the resistance as the magnetic field changes is shown in Fig. \ref{R vs T
in B}. In zero field, the resistance decreases with decreasing temperature
suggesting the existence of superconducting fluctuations. A magnetic field
destroys this downward curvature, and at some critical magnetic field, $%
B_{c} $, the resistance is independent of temperature. In magnetic fields
higher than $B_{c}$ the film is insulating, with $\partial R/\partial T<0$.
Figure \ref{R vs B} shows the resistance as a function of magnetic field for
different temperatures.

If the sheet resistance is normalized by the value of the critical
resistance at each thickness, $R/R_{c}(d),$ then all the data can be
collapsed onto a single curve. The collapse of the normalized resistance
data as a function of $\delta t$ for five samples is shown in Fig. \ref{B
collapse}. The critical exponent product $\nu z$, determined from the
temperature dependence of the parameter $t$ (inset of Fig.\ref{B collapse}.
), is found to be $\nu z=0.7\pm 0.2$, apparently independent of the film
thickness. The same exponent products were obtained using the alternative
method of plotting $(\partial R/\partial B)_{B_{c}}\ $vs. $T^{-1}$ on a
log-log plot and determining the slope which is equal to $1/\nu z$, as shown
on Fig. \ref{B exponent}.

Electric field scaling was also carried out for one of the samples.
Unfortunately, there was not enough data available for the insulating side
of the transition to carry out a complete analysis, but the data on the
superconducting side was sufficient to obtain the value of the critical
exponent product $\nu (z+1)$. The magnetic field dependence of the sheet
resistance for different values of electric field applied across the sample
is shown on Fig. \ref{R vs B in E}. The resistance data were then plotted as
a function of $(B-B_{c})$, and re-scaled by a parameter $t(E)$ to obtain the
best collapse of the data, shown in Fig. \ref{E collapse}. For the electric
field dependence of the parameter $t(E)$, shown in the inset of Fig. \ref{E
collapse}., the best power law fit was obtained for $\nu (z+1){\approx }1.4.$%
Combining this result with the result of the temperature scaling, it follows
that $z\approx 1$ and $\nu {{}\approx 0.7}$ for the magnetic field tuned
superconductor-insulator transition.

In contrast with our findings, previous studies of thin films of amorphous $%
InO_{x}$ \cite{Hebard} and $MoGe$ \cite{Yazdani} both showed $\nu \approx 1.3
$ and $z=1$ for the magnetic field tuned superconductor-insulator
transition. Our surprising result is also in obvious disagreement with the
prediction of the scaling theory (from which $\nu \geq 1$ \cite
{Chayes,Fisher B} for a disordered system), as well as with the
percolation-based models \cite{Shimshoni} (from which $\nu \approx 1.3$
would be expected).

\bigskip

\section{THICKNESS-TUNED SI TRANSITION}

For very thin films, the resistance increases exponentially with decreasing
temperature, while for the thicker films the resistance goes to zero as the
films become superconducting. At the critical thickness $d_{c},$ the
resistance is temperature independent, and the system is expected to stay
metallic down to $T=0$.

Using the same methods described above, the critical exponent product $\nu z$
was determined to be $1.2\pm 0.2$ when the superconductor-insulator
transition was tuned by changing the film thickness in zero magnetic field 
\cite{Markovic2}. A similar scaling behavior has been found in ultrathin
films of Bi by Liu et al.\cite{Liu}, with the critical exponent product $\nu
z=2.8$ on the insulating side and $\nu z=1.4$ on the superconducting side of
the transition. The fact that $\nu z$ was found to be different on the two
sides of the transition raises the question of whether the measurements
really probed the quantum critical regime. It is likely that the scaling was
carried out too deep into the insulating phase, forcing the scaling form
(Eq. \ref{T scaling}.) on films which were in a fundamentally different
insulating regime. Such films should not be expected to scale together with
the superconducting films, hence the discrepancy on the insulating side of
the transition. In the present work, the measurements were carried out at
lower temperatures than previously studied and with more detail in the range
of thicknesses close to the transition. Both sides of the transition scaled
with $\nu z{{}\approx }1.2$, which is close to the value obtained by Liu et
al. on the superconducting side of the transition. This result is also in
very good agreement with the predictions of the scaling theory \cite{Fisher
d}, renormalization group calculations \cite{Singh,Zhang,Herbut1}, and Monte
Carlo simulations \cite{Cha1,Wallin,Makivic}.

All previous experiments which studied the thickness or disorder tuned
superconductor-insulator transition were carried out in zero magnetic field.
An applied magnetic field is generally expected to change the universality
class of the transition since it breaks time reversal symmetry. One would
therefore expect the critical exponent product $\nu z$ to be different in
the presence of a finite magnetic field. Furthermore, the thickness-tuned
transition in a finite magnetic field might be expected to be in the same
universality class as the magnetic field-tuned transition at fixed thickness.

The thickness-tuned superconductor-insulator transition in a finite magnetic
field was probed by sorting the magnetoresistance data which were carefully
taken as a function of temperature and magnetic field for each film. A
detailed scaling analysis was carried out at fixed magnetic fields of: 0.5
kG, 1 kG, 2 kG, 3 kG, 4.5 kG and 7 kG for one set of films, and 12 kG for a
different set of films. For each value of the magnetic field, the resistance
was plotted as a function of the film thickness at different temperatures,
ranging from 0.14 K to 0.5 K, in order to determine the critical thickness
at that field. If the sheet resistance is normalized by the critical value
at each field $R/R_{c}(B)$, then the normalized resistance data as a
function of the scaling variable for all temperatures and all values of the
magnetic field collapsed onto a single curve, as shown on Fig. \ref{d
collapse}. The critical exponent product determined from the parameter $t(T)$
(Inset of Fig. \ref{d collapse}) was found to be $\nu z=1.4\pm 0.2$,
apparently independent of the magnetic field. Once again, the alternative
scaling procedure yielded very similar results, as shown in Fig. \ref{d
exponent}.

This value of the product $\nu z$ is a factor of two larger than that
obtained for the magnetic field-tuned transition. It is, however, very close
to that obtained from the analysis of the zero-field transition carried out
using data from the same set of films, which was $\nu z=1.2\pm 0.2.$ Given
the experimental uncertainties, it is hard to say whether this difference in
value of the exponent products reflects a difference between the
universality classes of the thickness driven transitions in zero and finite
magnetic field. These exponent products are close to those found in Monte
Carlo simulations of the (2+1)-dimensional classical XY model with disorder
by Cha and Girvin \cite{Cha1}.

\bigskip

\section{THE\ PHASE\ DIAGRAM}

Combining the data obtained from the thickness-tuned transitions in a fixed
magnetic field and the field-tuned transitions at the fixed thickness, one
can construct a phase diagram with thickness and magnetic field as
independent variables. This is shown in Fig.\ref{phase diagram}. The films
characterized by parameters which lie above the phase boundary are
''insulating'' ($\partial R/\partial T<0$ at finite temperatures), and the
ones below it are ''superconducting'' ($\partial R/\partial T>0$ at finite
temperatures). The phase boundary is a power law:

\begin{equation}
B_{c}\propto \left| d-d_{c}\right| ^{x}  \label{phase boundary}
\end{equation}
The best fit to the data yields $x=0.7$. Near the critical thickness for the
zero field transition, a simple dimensionality argument \cite{Fisher B}
suggests that the critical magnetic field should scale as:

\begin{equation}
B_{c}\propto \frac{\Phi _{0}}{\xi ^{2}}  \label{critical field}
\end{equation}
where $\Phi _{0}$ is the flux quantum. Since the correlation length is $\xi
\propto \left| d-d_{c}\right| ^{-\nu }$, one might expect the critical field
to be:

\begin{equation}
B_{c}\propto \left| d-d_{c}\right| ^{2\nu }  \label{theory phase boundary}
\end{equation}
According to the phase boundary obtained in this experiment (see Eq. \ref
{phase boundary}.), this would mean that $\nu =0.35$, a value not consistent
with the results of the scaling analysis carried out on the same films. It
also does not agree with $\nu =1.3$ obtained by Refs. \cite{Hebard} and \cite
{Yazdani}. There is no obvious physical reason for such a small value of $%
\nu $ and implied large values of $z$, so this discrepancy is a mystery at
this time. It is possible that the simple argument expressed in Eqs. \ref
{critical field} and \ref{theory phase boundary} is too naive.

Another surprising feature of the experimental results is that the critical
exponent product $\nu z$ evidently depends on whether the phase boundary is
crossed vertically (changing the thickness at a constant magnetic field), in
which case $\nu z\approx 1.4$, or horizontally (changing the magnetic field
at a fixed thickness), in which case $\nu _{B}z_{B}{{}\approx }\ 0.7$. One
might expect the critical exponents to not depend on the direction in which
the boundary is crossed. If, however, the actual tuning parameter were not
film thickness, but some other physical parameter which was a function of
thickness, a factor of two in the critical exponent product determined from
an analysis using thickness rather than the ''correct'' control parameter
might result. The ''correct'' control parameter might be some measure of
disorder, electron screening, damping, or Cooper pair density. The detailed
functional form of the thickness dependence of these parameters for
quench-condensed films is not known.

Another possibility is that there are actually two phase boundaries,
separating three different regimes, so that each exponent belongs to a
different phase boundary. There has been some indication of a vortex liquid
phase in between the superconducting (vortex glass) phase and the insulating
(Bose glass) phase \cite{Chervenak,vanOtterlo}. Since there only appears to
be one phase boundary, that is probably not the case. It is possible,
however, that the two boundaries could be indistinguishable over the range
of parameters explored in these studies, but would become apparent at higher
fields, greater film thicknesses, or lower temperatures. These matters need
to be investigated further.

\bigskip

\section{THE CRITICAL RESISTANCE}

The critical resistance for the field-driven transition, contrary to the
predictions of the dirty boson models, does not seem to be universal. Figure 
\ref{Rc vs B} shows that $R_{c}$ decreases as the critical field increases,
roughly in a linear fashion. Since thicker films have lower normal state
resistances and higher critical fields, this also means that $R_{c}$
decreases with increasing thickness and decreasing normal state resistance.
Very similar behavior was observed by Yazdani and Kapitulnik \cite{Yazdani}.
In order to explain the non-universal behavior of the critical resistance,
these authors proposed a two-channel conduction model, in which the
conductance due to the electron (fermion) channel adds to the conductance
due to the boson channel. When the unpaired electrons are strongly
localized, the conduction is mostly due to bosons, and the resistance is
close to $R_{Q}=h/4e^{2},$ as predicted by the boson Hubbard model. In the
opposite limit, unpaired electrons contribute significantly to the
conduction at the transition. Films with lower normal state resistances
would then have lower critical resistances due to the larger fraction of
normal electrons. The critical resistances in our experiment, however, are
all {\it greater} than $R_{Q}$ and their values could only be explained this
way if the quantum resistance due to pairs was itself greater than $R_{Q}$.

The conductance due to the electronic channel in a magnetic field might also
depend on the strength of the spin-orbit interactions, which is another
difference between our samples and those of Refs. \cite{Hebard} and \cite
{Yazdani}. The strength of the spin-orbit interactions is typically
proportional to $Z^{4}$, where $Z$ is the atomic number. Since $Bi$ is a
heavy metal, spin-orbit interactions are stronger than in the lighter $%
InO_{x}$ and $MoGe$. It is known that in the weakly localized systems with
strong spin-orbit interactions the magnetoresistance is positive, while it
is negative otherwise \cite{Bergmann,Lee}. If weakly localized unpaired
electrons really contributed significantly to the conduction at the magnetic
field-tuned superconductor-insulator transition at the experimentally
accessible finite temperatures, the contribution to the magnetoresistance
due to localization effects could have a positive or a negative sign,
depending on the strength of the spin-orbit interactions. This would make $%
R_{c}$ bigger in the case of Bi films, and smaller in the case of $InO_{x}$
and $MoGe$, consistent with experimental observations. There is, however, a
striking similarity in the magnetic field and normal state resistance
dependence of the critical resistance of the $Bi$ films and $MoGe$ films of
Ref. \cite{Yazdani}: even though their critical resistances fall on the
opposite sides of $R_{Q}$, they both decrease with magnetic field roughly
linearly, with almost the same slope.

Strictly speaking, the critical resistance is predicted to be universal only
at $T=0$, while the finite temperature corrections are expected to be scaled
by the with the Kosterlitz-Thouless transition temperature, $T_{c}$ \cite
{Fisher B}:

\begin{equation}
R_{c}(B_{c},T)=R_{c}^{*}\ +O(\frac{T}{T_{c}})^{2}  \label{Rc}
\end{equation}
where $R_{c}^{*}$ is the universal resistance at $T=0$, and $R_{c}$ is the
critical resistance at some finite temperature as measured in the
experiments. A closer look at the crossing plots such as that of Fig. \ref{R
vs B}, reveals that the critical resistance is indeed slightly temperature
dependent. A considerable amount of noise over the accessible temperature
range made it hard to compare this temperature dependence with Eq. \ref{Rc},
but qualitative behavior is shown on Fig. \ref{Rc vs T}. Normal state
resistances of the $MoGe$ films \cite{Yazdani} are a factor of 3-10 lower
than the $Bi$ films considered here, which means that our samples are
probing a different part of the phase diagram (normal state resistances are
inversely proportional to the film thickness in our experiment), and the
finite temperature corrections might be more important in one case then the
other. Indeed, somewhat higher critical resistances were found in $InO_{x}$
films if the temperature dependence of $R_{c}$ is taken into account \cite
{Gantmakher2}.

A recent analytical calculation of the critical resistance of a two
dimensional system at finite temperatures in the dirty boson model including
Coulomb interactions\cite{Herbut2} yielded a critical resistance of ${%
\approx }1.4R_{Q}$. The author suggested that the next order correction
would bring the result closer to $R_{Q}$. This result is in excellent
agreement with the critical resistance found in the present measurements,
which was $1.1-1.2R_{Q}$. Monte Carlo simulations of the $(2+1)$-dimensional
XY model without disorder \cite{Cha2} also find the critical resistance to
be $R_{c}=7.7k\Omega $, again very close to the value found in this work.

\bigskip

\section{DISCUSSION}

A lot of attention has been focused recently on the effects of dissipation
on SI transitions \cite
{Shimshoni,Mason,Wagenblast,Kramer,Chakravarty,Rimberg}. Within the picture
proposed by Shimshoni {\it et al}.\cite{Shimshoni}, the transition between
the superconducting and the insulating state is of a percolative nature. On
the insulating side of the transition, electrical transport occurs through
activation or tunneling of Cooper pairs between superconducting domains.
Likewise, on the superconducting side, vortices tunnel from one insulating
domain to another. Using incoherent Boltzmann transport theory, Shimshoni 
{\it et al}. derive resistivity laws in different temperature regimes and
predict finite dissipation at $T=0$ for all values of the magnetic field.
Their results seem to be supported by measurements on several different
systems: thin films \cite{Ephron,Yazdani}, 2D Josephson junction arrays \cite
{vdZant}, Si MOSFETs \cite{Kravchenko} and QH systems \cite{Shahar}, where a
saturation of the resistance at low temperatures is observed and attributed
to dissipation effects. The percolative nature of the transition can explain
the value of $\nu \approx 1.3$ found in most of the field-tuned experiments
on thin films \cite{Hebard,Yazdani}, as well as the apparent symmetry
between insulating and the conducting phase observed in other experiments 
\cite{vdZant,Kravchenko,Simonian,Shahar}.

In contrast with the above mentioned results, we do not observe any
saturation in the temperature dependence of the resistance as the
temperature decreases, or in other words, $\delta R/\delta T$ is non-zero
down to the lowest temperatures, which were above 0.1K. Of course
investigation down to even lower temperatures might lead to a different
conclusion. However a satisfactory fit to the resistivity laws predicted by
Shimshoni {\it et al.}\cite{Shimshoni} could not be obtained.

Mason and Kapitulnik \cite{Mason} recently proposed a new phase diagram for
the SI transition which takes into account the possibility of a coupling of
the system to a dissipative bath. They argued that this coupling, which
becomes important when the critical point is approached, can result in a
new, metallic-like phase. In this picture, a direct SI transition is still
possible for very weak coupling, while for a stronger coupling the system
goes through a metallic phase and is truly superconducting only at the
lowest magnetic fields.

The fact that the typical sheet resistances of our samples are about a
factor of five higher than those in which resistance leveling was observed 
\cite{Ephron} might just mean that our samples are in the weak coupling
regime. However, the correlation length exponent determined in our
experiment for the magnetic field-tuned transition, using two different
methods, on different physical samples and at several levels of disorder was
found to be $\nu \approx 0.7$,which is not consistent with the exponent
expected from the classical 2D percolation theory, $\nu =4/3,$ even with
much more generous error bars.

A coherence length exponent of 0.7 is also inconsistent with what was
believed to be an exact theorem, \cite{Chayes} which predicts $\nu \geq 1$
in two dimensions in a presence of disorder. It is interesting to note that
our exponent agrees with the result of the classical 3D XY model which is
suggested to be relevant in the absence of disorder \cite{Fisher g}.
Numerical simulations of a (2+1)-dimensional XY model \cite{Cha2} and the
Boson-Hubbard model at $T=0$ \cite{Kisker} without disorder also find $z=1$
and $\nu =0.7.$ However, recently it was suggested that the nature of
disorder averaging may introduce a new correlation length, different from
the intrinsic one, which might lead to $\nu <1$ even for a disordered system 
\cite{Pazmandi}.

There is also a possibility that the local dissipation coupled to the phase
of the superconducting order parameter due to gapless electronic excitations
might change the universality class of the system and lead to a
non-universal critical resistance \cite{Wagenblast}. The critical resistance
would then be expected to increase with increasing damping due to
dissipation. The latter would be expected to increase with decreasing normal
state resistance. However, we observe that the critical resistance decreases
as the normal state resistance decreases, which is exactly the {\it opposite 
}of the behavior predicted by Wagenblast {\it et al}.

We should note that $\nu \approx 0.7$ was also found for the magnetic
field-tuned insulator-conductor transition in Si MOSFET samples \cite
{Phillips}, suggesting a possible connection between the two phenomena.

Our results for the magnetic field-tuned SI transition seem to be consistent
with the predictions of bosonic models, rather than percolation models. This
is further supported by the transport studies in the insulating regime,
where the magnetoresistance cannot be explained by the weak localization
theory only \cite{Markovic1}, and the temperature dependence of the
resistance fits the predictions of Das and Doniach \cite{Das2} for the
bosonic conduction. These observations still need to be reconciled with the
results of the tunneling experiments which find no superconducting gap in
the insulating regime. The tunneling experiments might however be
emphasizing regions of the samples containing quasi-localized single
electron states below the gap, or those in which the amplitude fluctuations
break the system into superconducting ''islands'' with finite spectral gaps
in the density of states, as recently predicted \cite{Ghosal}. A highly
non-uniform gap has also been predicted by Herbut \cite{Herbut3} for the
case of large disorder. This problem might be clarified using spatially
resolved scanning tunneling spectroscopy at low temperatures, which may be
able to detect{\it \ local} variations in the density of states.

Such studies might also help answer the question as to why $\nu $ different
for the thickness- and magnetic field-tuned transitions on the same samples.
In the case of the thickness-tuned transition, the correlation length
exponent is close to what might be expected from the percolation theory.
There is a major difference between the magnetic field-tuned and the
thickness-tuned transitions: when the transition is tuned by the magnetic
field, the microstructure of the sample stays fixed, while in the case of
the thickness-tuned transitions it changes slightly with each film in the
sequence. It may be that in this case the percolation effects become
relevant, complicating the determination of the critical exponents \cite
{Sheshadri}.

Finally, the shape of the phase boundary poses a further challenge to
theorists. We are currently investigating the role of the dissipation in
this system in more detail, using a 2D electron gas as a substrate, similar
to the experiment of Rimberg {\it et al}. \cite{Rimberg}.

We gratefully acknowledge useful discussions with A. P. Young, S. Sachdev
and P. Phillips. This work was supported in part by the National Science
Foundation under Grant No. NSF/DMR-9623477. 

\begin{figure}[tbp]
\caption{Resistance per square as a function of temperature in different
magnetic fields, ranging from 0kG (bottom) to 12kG (top), with 1kG
increments.}
\label{R vs T in B}
\end{figure}

\begin{figure}[tbp]
\caption{Resistance per square as a function of magnetic field for a bismuth
film close to the transition. Different curves represent different
temperatures: 0.15, 0.17, 0.19, 0.2, 0.25, 0.3 and 0.35K.}
\label{R vs B}
\end{figure}

\begin{figure}[tbp]
\caption{Normalized resistance per square as a function of the scaling
variable, $T^{-1/\nu z}|B-B_{c}|$. Each symbol represents one film at
different temperatures (only a small portion of the data is shown for
clarity). Inset: The fitting a power law to the temperature dependence of
the parameter t determines the value of $\nu z.$}
\label{B collapse}
\end{figure}

\begin{figure}[tbp]
\caption{The critical exponent product $\nu z$ for the magnetic field-tuned
transition as determined by the inverse slope of $\partial R/\partial B$ at
the critical value of $B_{c}$ plotted vs $1/T$.}
\label{B exponent}
\end{figure}

\begin{figure}[tbp]
\caption{Resistance per square as a function of magnetic field at different
electric fields across the film: 0.5 (botttom), 1.0, 1.5, 2.0, 2.5, 3.0 and
3.5 V/m (top). Only $B<B_{c}$ is shown where the resistance increases with
increasing electric field. The temperature is 0.7K.}
\label{R vs B in E}
\end{figure}

\begin{figure}[tbp]
\caption{Resistance per square as a function of the scaling variable, $%
t|B-B_{c}|$, for different electric fields: 0.5 , 1.0, 1.5, 2.0, 2.5, 3.0
and 3.5 V/m. Here $t=E^{-1/\nu (z+1)}$ is treated as an adjustable parameter
to obtain the best collapse of the data. Inset: The fitting a power law to
the temperature dependence of the parameter t determines the value of $\nu
z. $}
\label{E collapse}
\end{figure}

\begin{figure}[tbp]
\caption{Normalized resistance per square as a function of the scaling
variable, $t|d-d_{c}|$, in different magnetic fields: 0.5\ (squares), 1.0\
(circles), 3.0\ (crosses), 4.5\ (triangles) and 7.0 kG\ (diamonds). Inset:
The fitting a power law to the temperature dependence of the parameter t
determines the value of $\nu z.$}
\label{d collapse}
\end{figure}

\begin{figure}[tbp]
\caption{The critical exponent product $\nu z$ for the thickness-tuned
transition as determined by the inverse slope of $\partial R/\partial d$ at
the critical value of $d_{c}$ plotted vs $1/T$.}
\label{d exponent}
\end{figure}

\begin{figure}[tbp]
\caption{The phase diagram in the d-B plane in the T=0 limit. The points on
the phase boundary were obtained from thickness tuned transitions
(triangles) and magnetic field-tuned transitions (circles). The solid line
is a power law fit. Here d$_{c}$ is taken to be the critical thickness in
zero field.}
\label{phase diagram}
\end{figure}

\begin{figure}[tbp]
\caption{The critical resistance as a function of the critical field for a
series of bismuth films. Here $R_{c}$ decreases with increasing thickness,
as thicker films have lower normal state resistances and higher critical
fields.}
\label{Rc vs B}
\end{figure}
\begin{figure}[tbp]
\caption{The critical resistance as a function of temperature for a 12.353
\AA\ thick film.}
\label{Rc vs T}
\end{figure}
command. 
command. 

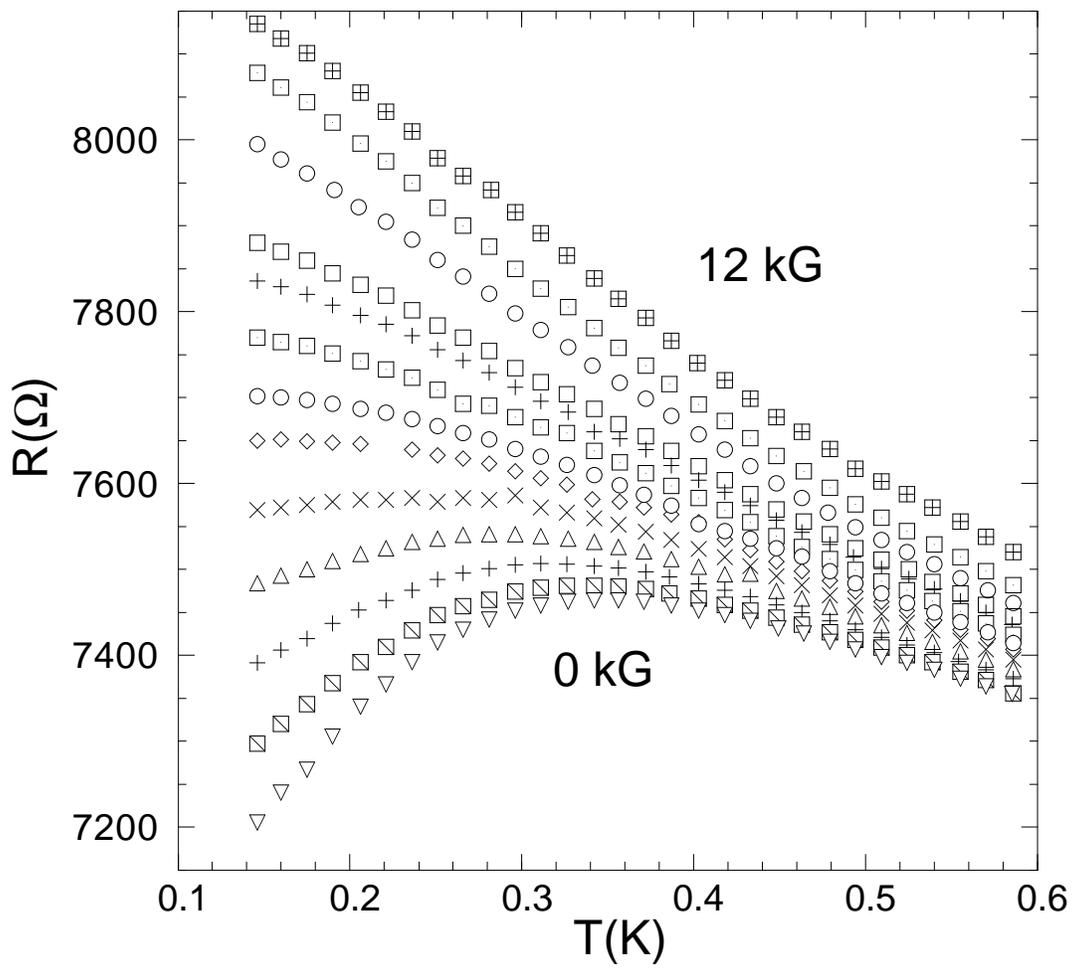
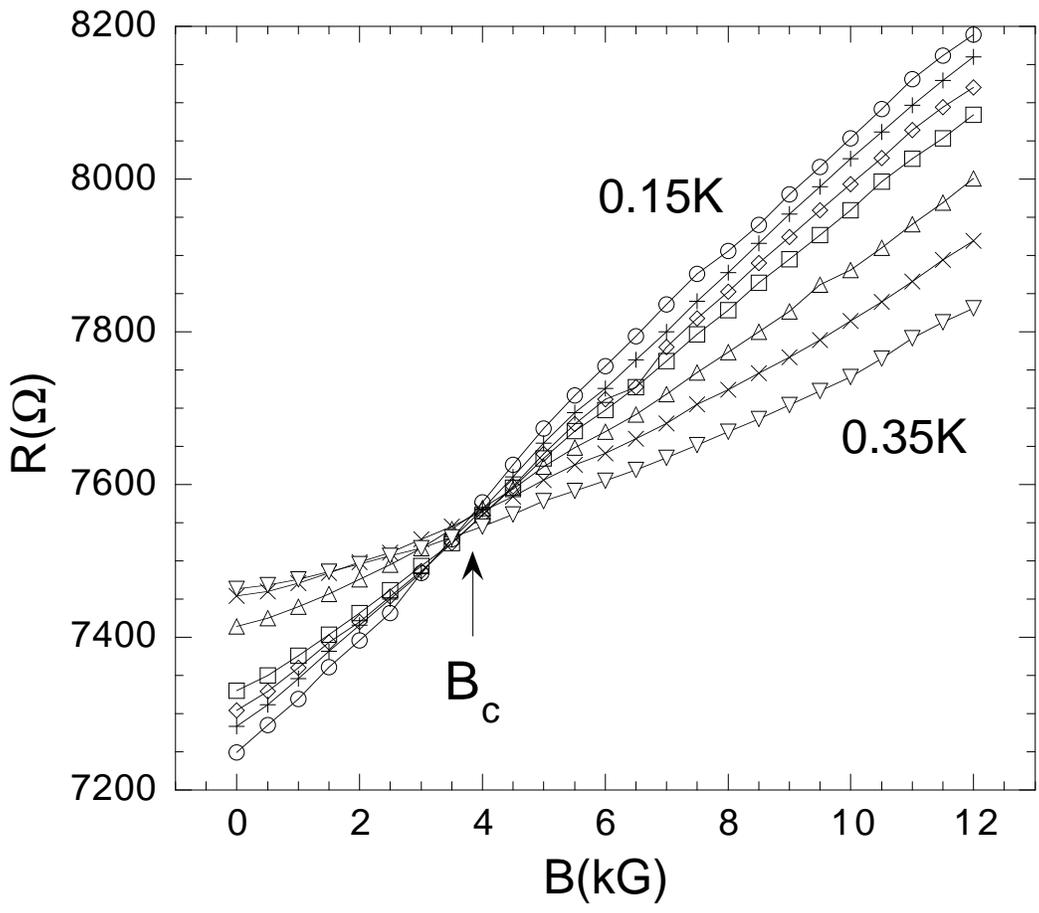
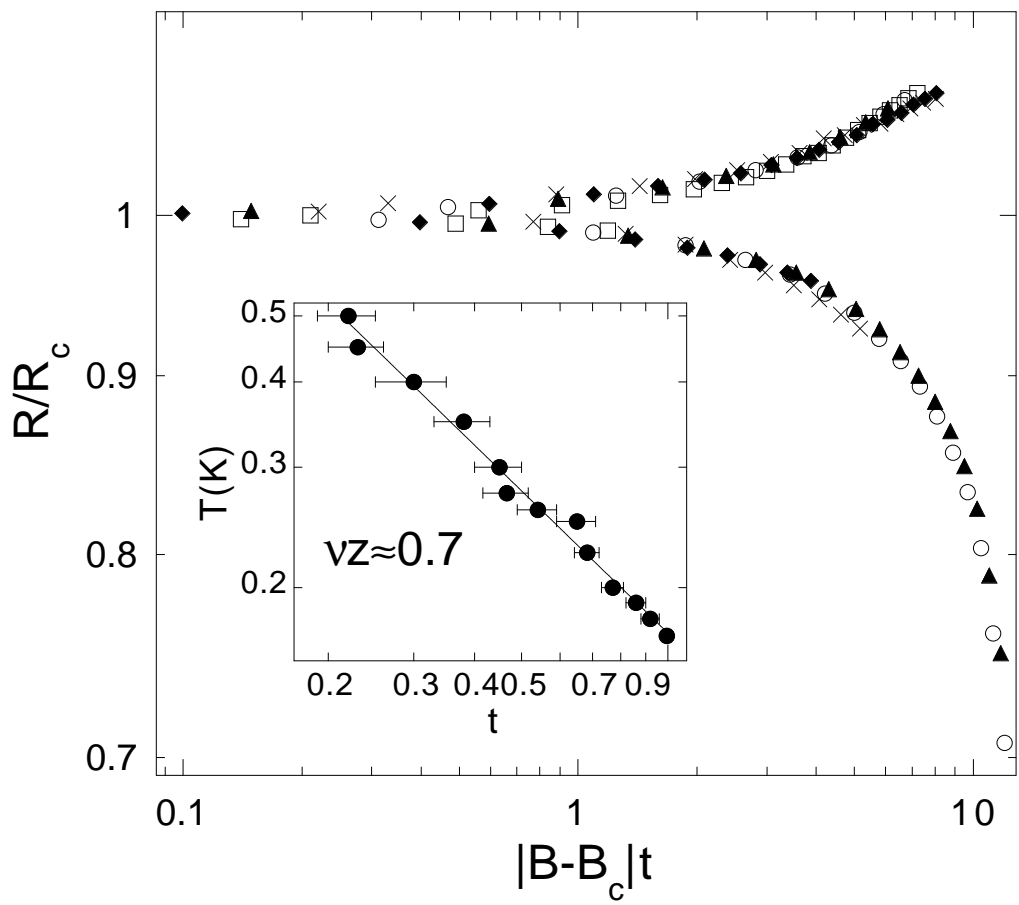
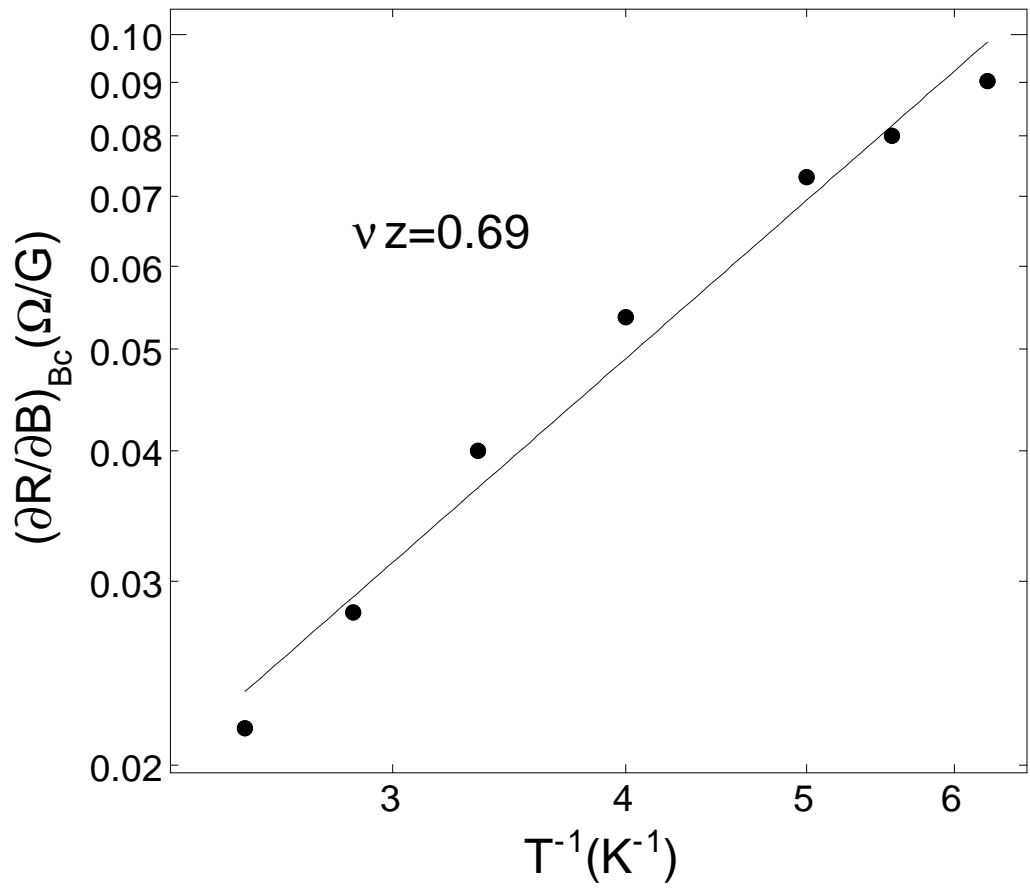
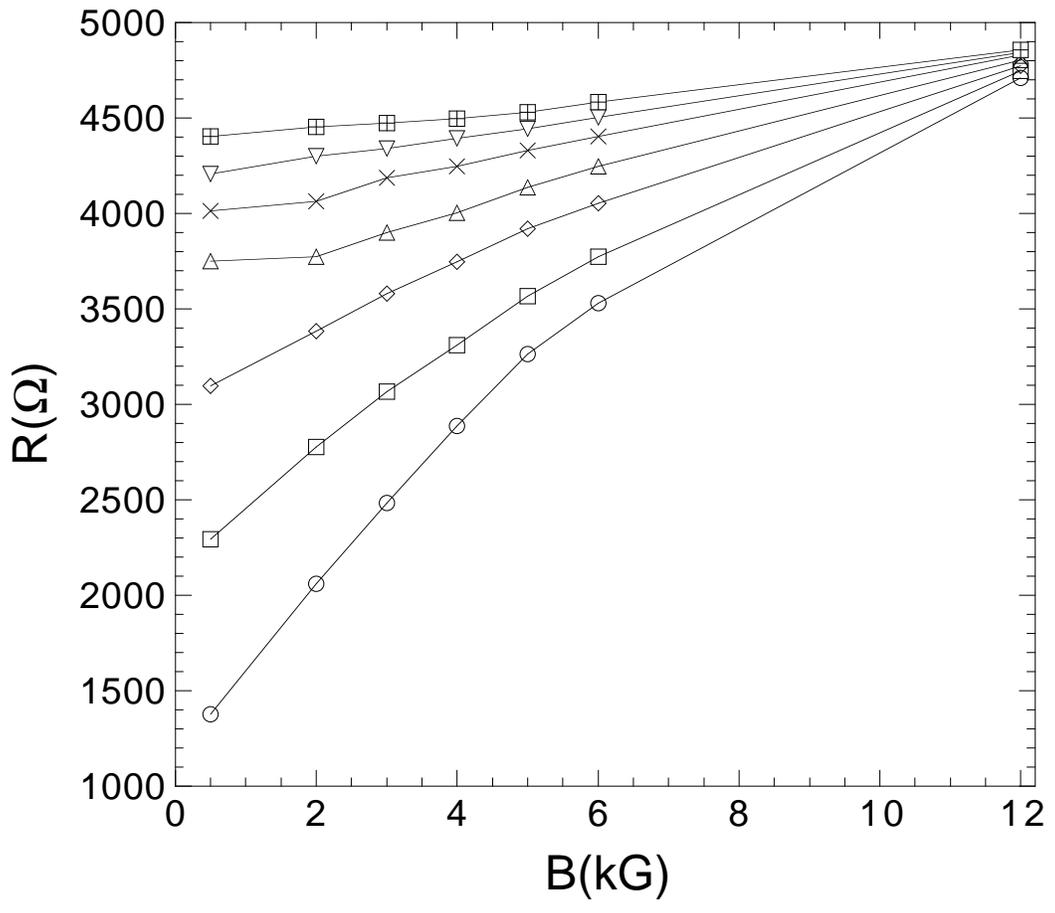
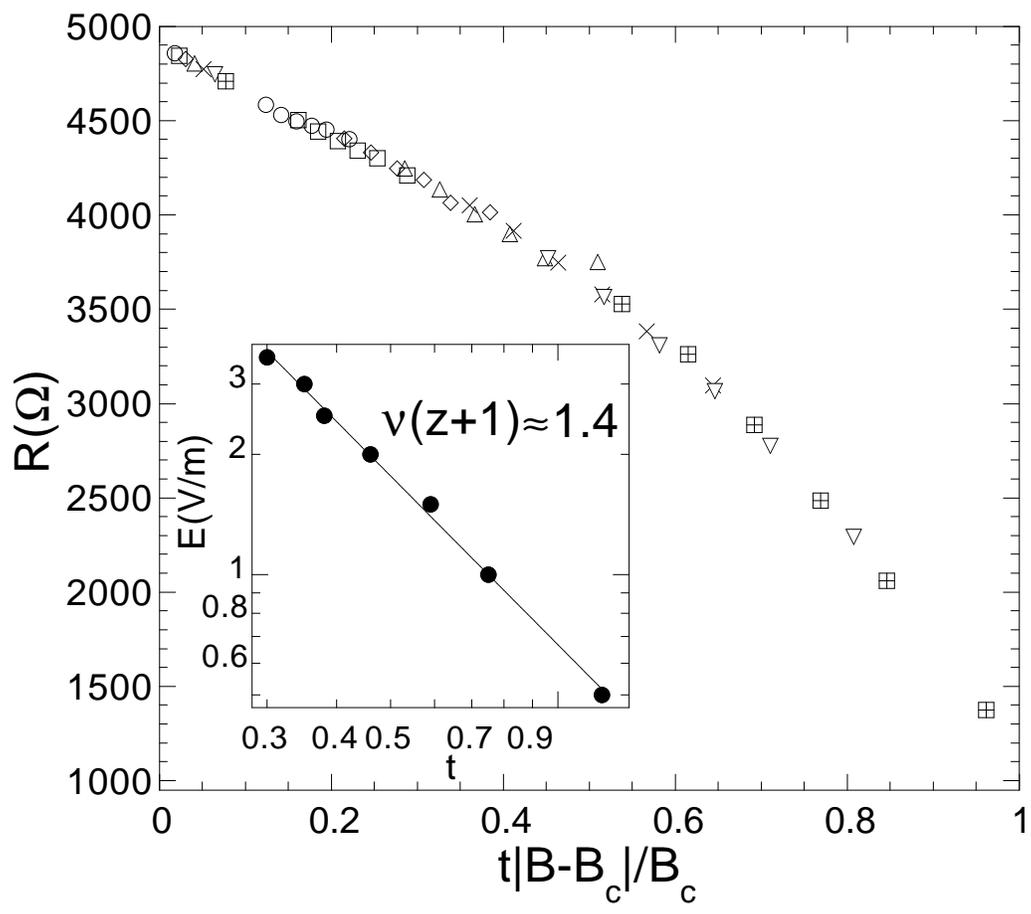
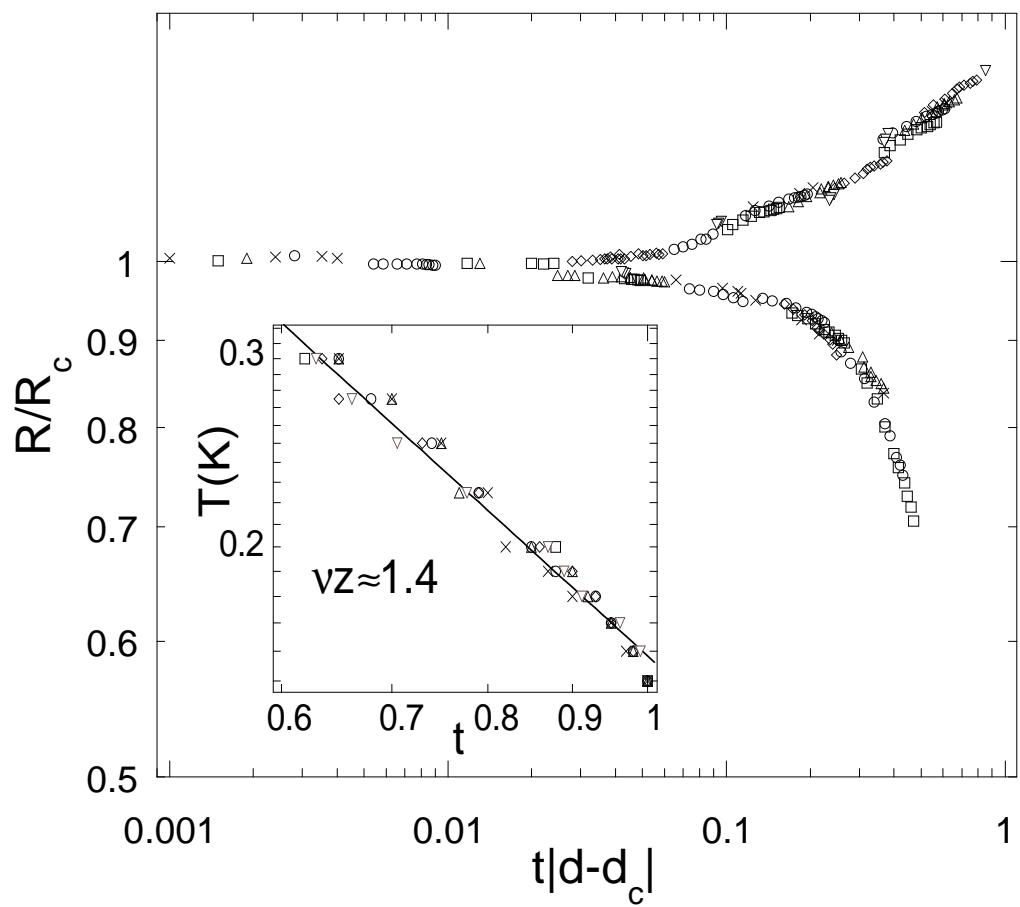
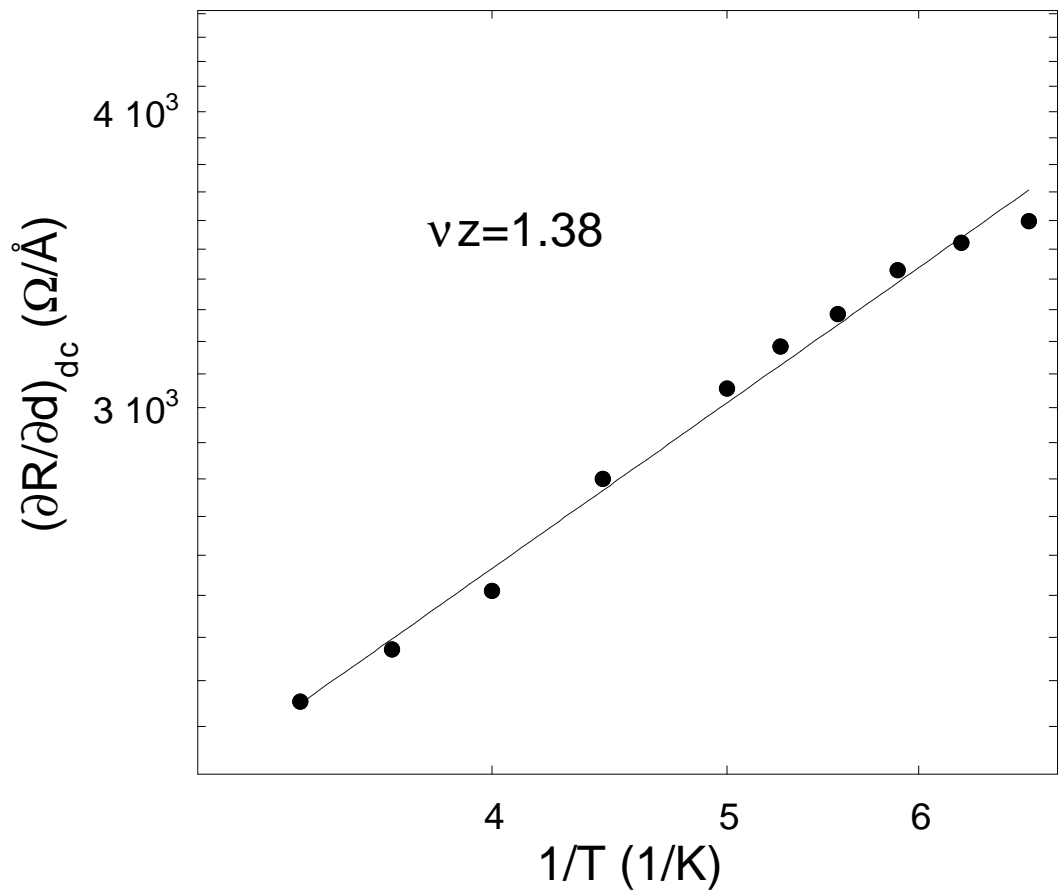
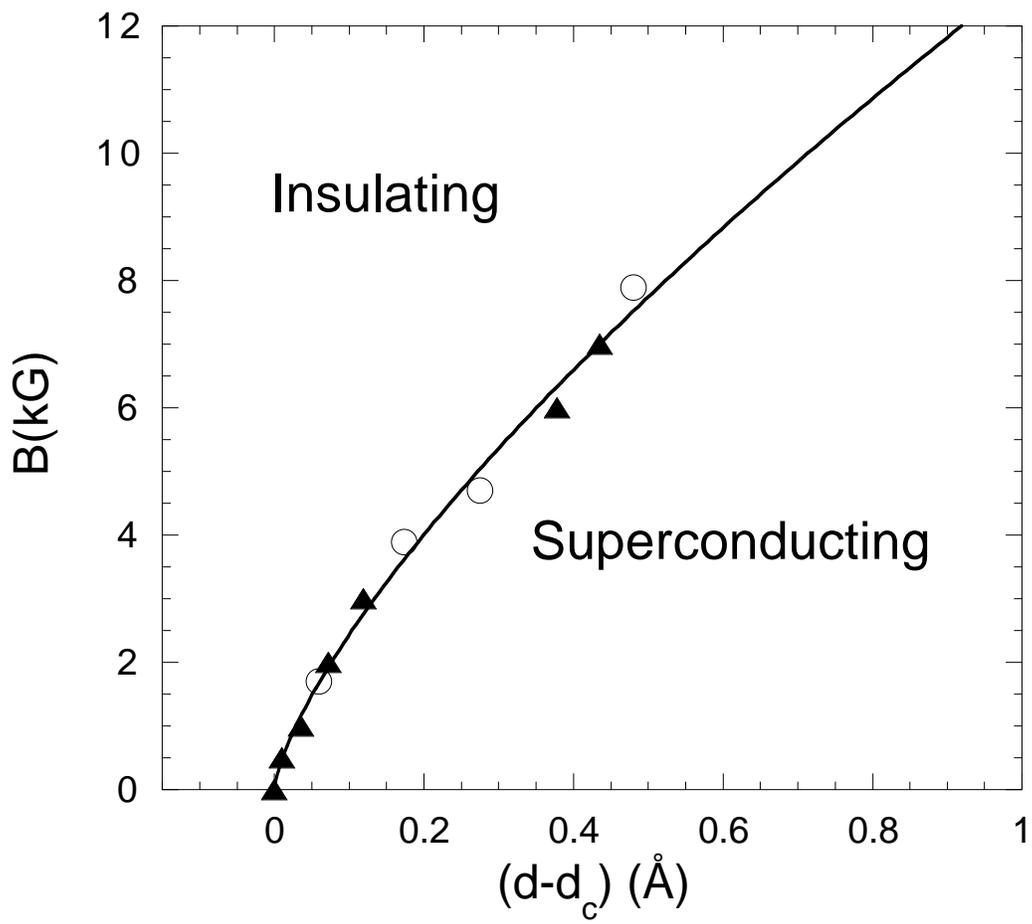
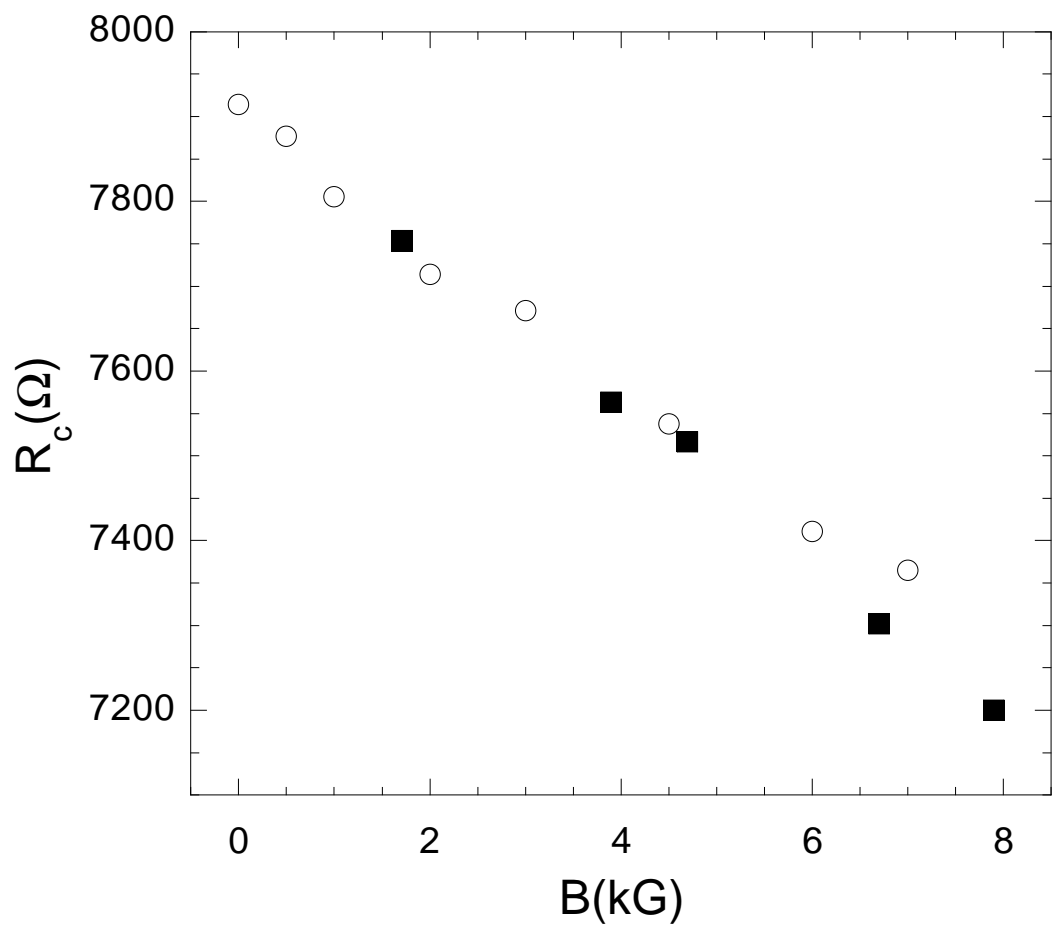
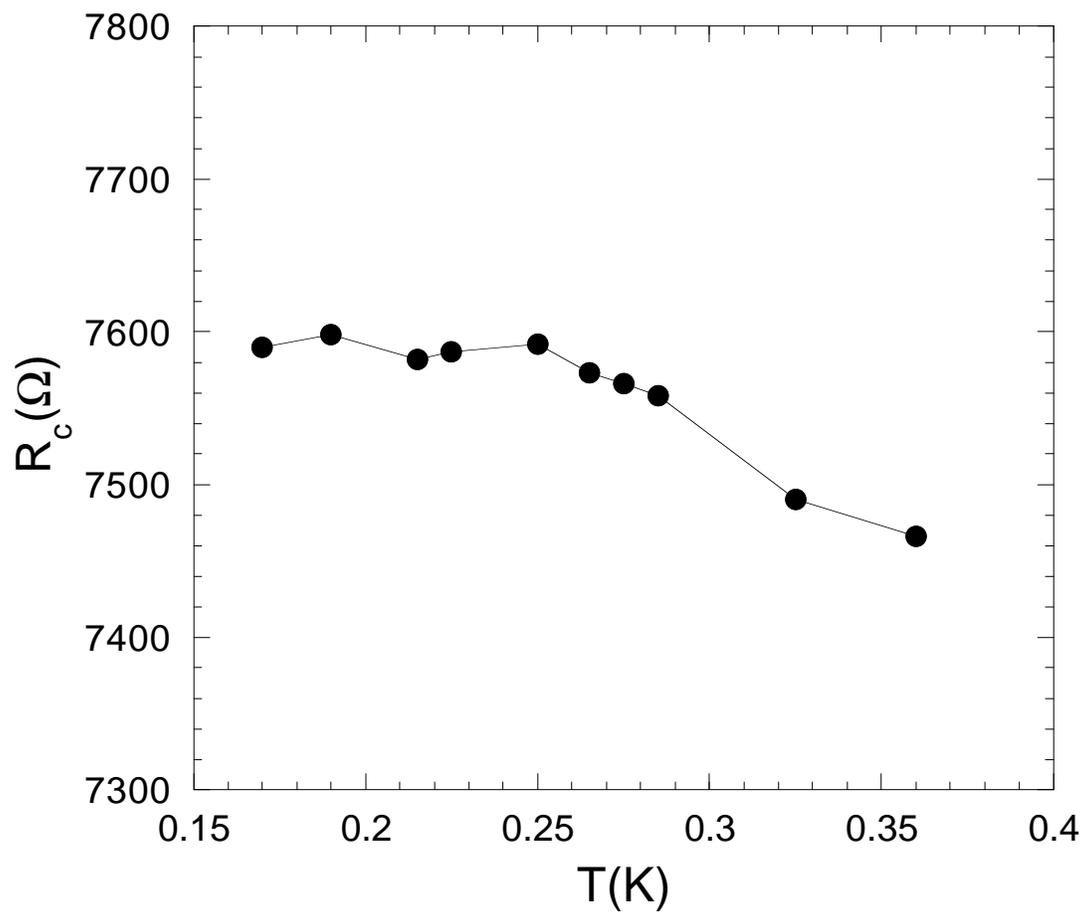
\end{document}